\documentclass[aip,reprint]{revtex4-1}

\usepackage[english]{babel}
\usepackage{braket}
\usepackage{amsthm}

\usepackage{mathtools}
\usepackage{siunitx}
\usepackage{accents}
\usepackage{natbib}
\usepackage{xcolor}

\usepackage{graphicx}
\usepackage{dcolumn}
\usepackage{bm}

\usepackage[utf8]{inputenc}
\usepackage[T1]{fontenc}
\usepackage{mathptmx}
\usepackage{etoolbox}

\usepackage[]{xcolor}
\usepackage{soul}
\usepackage{cancel}

\DeclareSIUnit\angstrom{\text {Å}}

\draft 

\begin{document}


\title{Coupled cluster cavity Born-Oppenheimer approximation for electronic strong coupling} 
%


\author{Sara Angelico}
\affiliation{Department of Chemistry, Norwegian University of Science and Technology, 7491 Trondheim, Norway}

\author{Tor S. Haugland}
\affiliation{Department of Chemistry, Norwegian University of Science and Technology, 7491 Trondheim, Norway}

\author{Enrico Ronca}
\affiliation{Dipartimento di Chimica, Biologia e Biotecnologie, Università degli Studi di Perugia, Via Elce di Sotto, 8, 06123, Perugia, Italy}

\author{Henrik Koch}
\email[]{henrik.koch@ntnu.no}
\affiliation{Department of Chemistry, Norwegian University of Science and Technology, 7491 Trondheim, Norway}
\affiliation{Scuola Normale Superiore, Piazza dei Cavalieri 7, 56126 Pisa, Italy}


\date{\today}

\begin{abstract}
Chemical and photochemical reactivity, as well as supramolecular organization and several other molecular properties, can be modified by strong interactions between light and matter. Theoretical studies of these phenomena require the separation of the Schr\"odinger equation into different degrees of freedom as in the Born-Oppenheimer approximation. In this paper, we analyze the electron-photon Hamiltonian within the cavity Born-Oppenheimer approximation (CBOA), where the electronic problem is solved for fixed nuclear positions and photonic parameters. Specifically, we focus on intermolecular interactions in representative dimer complexes. The CBOA potential energy surfaces are compared with those obtained using a polaritonic approach, where the photonic and electronic degrees of freedom are treated at the same level. This allows us to assess the role of electron-photon correlation and the accuracy of CBOA. 
\end{abstract}


\maketitle 

\section{Introduction}
\noindent
During the past few years, increasing attention has been devoted to the possibility of exploiting strong light-matter interactions to modify molecular properties. Several recent studies have shown that under proper conditions electromagnetic fields can affect supramolecular organization, \cite{joseph2021supramolecular,nagarajan2021chemistry,hirai2021selective,kulangara2022manipulating} optical properties \cite{schwartz2013polariton,kadyan2021boosting,george2015ultra} and photochemical processes. \cite{climent2022not,takahashi2019singlet,gu2021optical,fregoni2018manipulating,fregoni2020strong} Moreover, different experimental works demonstrated the possibility to modify ground state reactivity by means of strong interactions between vibrational states and a cavity field. \cite{thomas2019tilting,lather2019cavity,thomas2016ground,sau2021modifying,hirai2020modulation,ahn2023modification} \\ 
The most common experimental setups to reach strong coupling conditions are Fabry-Pérot cavities \cite{nagarajan2021chemistry,perot1899application}. 
\begin{figure}%
    \centering%
    \includegraphics[width=.9\columnwidth]{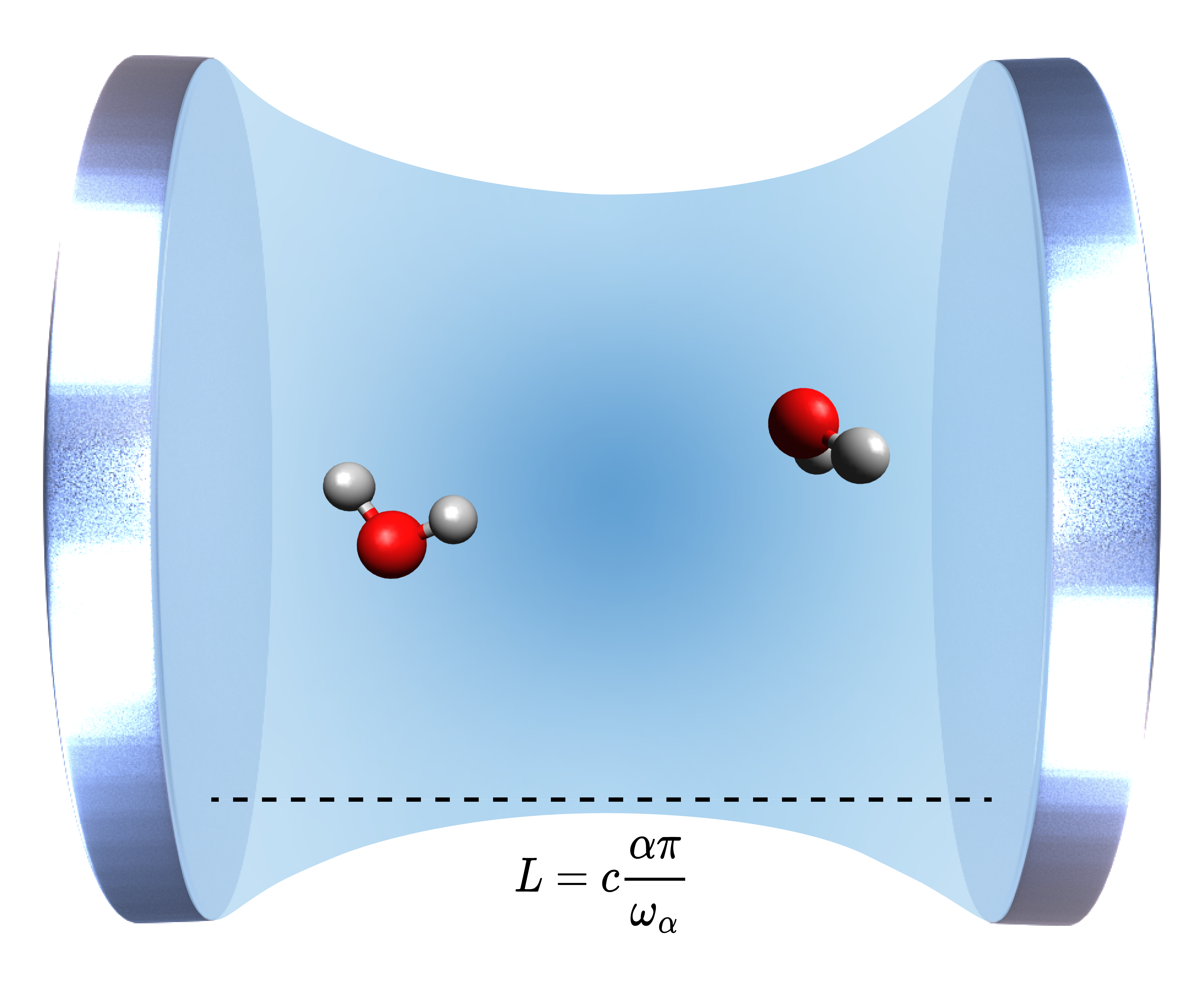}%
    \caption{Example of a Fabry-Pérot cavity. The relation between the length of the cavity L and the frequency of its modes $\omega_\alpha$ is reported.}%
    \label{fig:cavity}%
\end{figure}%
These devices can be schematized as two parallel highly reflecting mirrors surrounding the molecular system and separated by a very short distance (See Fig.~\ref{fig:cavity}). In these conditions, the electromagnetic field interferes with itself generating standing waves whose frequencies are related to the distance between the cavity mirrors. When the exchange of energy between the field and the molecules is faster than any decay process, strong light-matter interactions occur. Depending on the frequency of the electromagnetic field, that can be in resonance with either electronic or vibrational excitations, it is possible to reach either the electronic (ESC) or vibrational (VSC) strong coupling regimes, respectively.  \\
When strong light-matter interactions occur, the electromagnetic field effects cannot be included perturbatively, and they require an explicit quantum treatment. The introduction of the quantized electromagnetic field in the Hamiltonian, nevertheless, leads to an increased dimensionality of the problem, that now depends on electronic, nuclear and photonic degrees of freedom. As a consequence, the complete Schr\"odinger equation has to be simplified for these systems. This can be done generalizing the Born-Oppenheimer approximation, where the electronic and nuclear coordinates are separated to generate two problems with lower dimensionality.
Nevertheless, for polaritonic systems, the division of the problem is not straightforward. Up to now, two main approaches to generalize the Born-Oppenheimer approximation to this framework have been proposed. In the first case, the so-called polaritonic approach treats the photonic degrees of freedom at the same level as the electrons, while considering fixed nuclear configurations. \cite{qed-cc,fregoni2020photochemistry} The second approach, developed by Flick et al., \cite{cboa,cboa_weaktostrong} is the cavity Born-Oppenheimer approximation (CBOA), which focuses on the formation of vibrational polaritonic states, thus separating the electronic coordinates from the photonic and nuclear ones. \cite{cboa,cboa_weaktostrong} These two methods provide complementary points of view on coupled electron-nuclear-photon systems and have been applied to the study of different cases, ranging from molecular properties \cite{haugland2021intermolecular,riso2022characteristic,castagnola2023polaritonic,philbin2023molecular,liebenthal2023assessing} to chemical and photochemical reactivity. \cite{fregoni2018manipulating,kim2012control,schafer2022shining,fregoni2020photochemistry,Schnappinger2023Nonadiabatic} Both approaches, however, require the introduction of methods capable of accurately describe strong electron-photon or nuclear-photon interactions. \\
As far as electron-photon correlation is concerned, during the past few years some electronic structure methods have been generalized to the polaritonic framework, including the quantized electromagnetic field. Some recent methods include quantum electrodynamics Hartree-Fock (QED-HF)\cite{qed-cc}, quantum electrodynamics coupled cluster (QED-CC)\cite{qed-cc,haugland2021intermolecular}, quantum electrodynamics full configuration interaction (QED-FCI) \cite{qed-cc,haugland2021intermolecular} and quantum electrodynamics density functional theory (QEDFT). \cite{qedft1,qedft2,qedft3} \\
In this paper, we formulate electronic structure methods within the cavity Born-Oppenheimer approximation. We report benchmark results for Hartree-Fock (CBO-HF \cite{schnappinger2023cavitybornoppenheimer}), complete active space configuration interaction (CBO-CASCI), full configuration interaction (CBO-FCI), coupled cluster theory with perturbative double excitations (CBO-CC2), coupled cluster with single and double excitations (CBO-CCSD), coupled cluster with perturbative triple excitations (CBO-CC3) and coupled cluster with single, double and triple excitations (CBO-CCSDT). We then specifically focus on intermolecular interactions in three selected dimers that are representative examples of van der Waals forces, hydrogen bonding, and dipole-induced dipole interactions using CBO-CCSD. Comparing these results to the QED-CCSD approach, we analyze the flexibility of the cavity Born-Oppenheimer framework and the role of electron-photon correlation. \\
This paper is organized as follows. In Sec.~\ref{Sec:CBOA}, we give a brief introduction to the cavity Born-Oppenheimer approximation, followed in Sec.~\ref{Sec:Reformulation of electronic structure methods to CBOA} by the introduction of post Hartree-Fock methods within CBOA. In Sec.~\ref{Sec:Benchmark} we discuss the benchmark results and in Sec.~\ref{Sec:Intermolecular interactions in dimers} we present the study of intermolecular interactions. Our final remarks are given in Sec.~\ref{Sec:Conclusions}.
\section{Cavity Born-Oppenheimer approximation}
\label{Sec:CBOA}
\noindent
In the description of strongly coupled systems, we must include the quantized electromagnetic field and its interactions with matter. To this end, the system is usually described in Coulomb gauge within the dipole approximation and the Power-Zienau-Woolley (PZW) framework \cite{ruggenthaler2022understanding,weight2023theory,cboa_weaktostrong}
\begin{align}
\label{eq:CBOA-Ham-Theory}
     H &= H_e + T_N \nonumber \\
     &+ \sum_\alpha\bigg(\frac12 \big(\hat{p}_\alpha^2 +  \omega_\alpha^2 \hat{q}_\alpha^2 \big)+ \omega_\alpha \hat{q}_\alpha (\bm{\lambda}_\alpha \cdot \mathbf{d}) + \frac12  (\bm{\lambda}_\alpha \cdot \mathbf{d})^2 \bigg).
\end{align}
This Hamiltonian includes, besides the usual electronic Hamiltonian H$_e$ and the kinetic energy of the nuclei T$_N$, additional terms, related to the quantized electromagnetic field. In particular, the electromagnetic field is expressed as the sum of harmonic oscillators with frequency $\omega_\alpha$ described in terms of the displacement operator $\hat{q}_\alpha$ and the momentum operator $\hat{p}_\alpha$. Moreover, the fourth and the fifth terms in Eq.~\eqref{eq:CBOA-Ham-Theory} describe the light-matter interactions. In the dipole approximation, these are mediated by the dipole moment operator of the molecule $\mathbf{d}$ and the coupling strength $\bm{\lambda}_\alpha= \sqrt{\frac{4\pi}{ V_\alpha}}\bm{\epsilon}_\alpha$, that depends on the $\alpha$-mode quantization volume $V_\alpha$ and on the polarization of the field $\bm{\epsilon}_\alpha$. 
The last term is the dipole self-energy, that describes how the polarization of matter acts back on the photon field.\cite{cboa_weaktostrong,cboa} In addition, this term ensures that the Hamiltonian is bounded from below, which ensures a ground state for the coupled light-matter system. \cite{no_gs_dse, schafer2018ab} The displacement and momentum operators can be expressed in terms of the photonic creation and annihilation operators as
\begin{equation}
    \hat{q}_\alpha = \frac{1}{\sqrt{2\omega_\alpha}}(b^\dagger_\alpha + b_\alpha)
\end{equation}
and
\begin{equation}
     \hat{p}_\alpha=i\sqrt{\frac{\omega_\alpha}{2}}(b^\dagger_\alpha - b_\alpha) .
\end{equation}
Note that the displacement operator $\hat{q}_\alpha$ is proportional to the electric displacement field operator $\mathbf{\hat{D}_\alpha} =\frac{1}{4\pi} \omega_\alpha \bm{\lambda}_\alpha \hat{q}_\alpha$, while $\hat{p}_\alpha$ is proportional to the magnetic field. \cite{cboa,schnappinger2023cavitybornoppenheimer}   \\ \\
In the cavity Born-Oppenheimer approximation, the solution of the Schr\"odinger equation is divided in two parts. First, an electronic wave function $\psi_k(\mathbf{r};\mathbf{R},\mathbf{q})$ is determined for a fixed nuclear and photonic configuration treating the corresponding displacement coordinates as parameters. Subsequently, a vibropolaritonic wave function $\chi_{kv}(\mathbf{R},\mathbf{q})$ that explicitly depends on both nuclear and photonic degrees of freedom is determined. Here, $\nu$ labels the vibropolaritonic states. Finally, a stationary state is approximated as the product of these two terms
\begin{equation}
    \Psi_{kv}(\mathbf{r},\mathbf{R},\mathbf{q}) = \psi_k(\mathbf{r};\mathbf{R},\mathbf{q})\chi_{kv}(\mathbf{R},\mathbf{q}) .
\end{equation}
Using the terminology commonly used in the Born-Oppenheimer approximation, the photonic degrees of freedom are here considered "slow" and possibly resonant with the nuclear ones. \cite{fregoni2020photochemistry} Analyzing this in more detail, since the photonic momentum operator $\hat{p}_\alpha$ is proportional to the magnetic field, neglecting it corresponds to assuming that the magnetic field is small. \cite{cboa} This is in line with the dipole approximation, in which the effects of the magnetic field are disregarded. Moreover, since $p_\alpha = \frac{d}{dt}q_\alpha$, it follows that the displacement field changes slowly over time. Overall, the cavity Born-Oppenheimer approximation is expected to be valid when the electrons are able to readily adapt to the slow variations of the displacement field. \cite{cboa,flick2018strong} A more rigorous and quantitative assessment of the validity of this approximation, nevertheless, would require the computation of non-adiabatic coupling elements. For this reason, as in standard Born-Oppenheimer approximation, this approach should be accurate when the potential energy surfaces are well separated from each other. Overall, we expect the cavity Born-Oppenheimer approximation to provide a useful framework for the approximate treatment of the effects of the field on the electronic subsystem, and on the electronic ground state in particular, in vibrational strong coupling conditions. \cite{cboa_weaktostrong,PhysRevX.9.021057,Schnappinger2023Nonadiabatic} As for the ESC regimes, where electron-photon correlation is particularly relevant, we expect this approximation to only partially reproduce the effects of the electromagnetic field.
\section{Electronic structure methods within CBOA}
\label{Sec:Reformulation of electronic structure methods to CBOA}
\noindent
As already mentioned, in the cavity Born-Oppenheimer approach a stationary state is approximated as the product of an electronic and a vibropolaritonic wave function. In the following, we will focus on the electronic problem. \\
Here, the nuclear and photonic degrees of freedom are kept fixed, and the terms describing the corresponding kinetic energies are neglected. The electronic CBOA Hamiltonian, then, is the sum of the usual electronic Hamiltonian, the potential energy of the field, and the field interactions with matter 
\begin{equation}
\label{ham_cboa_orig}
   H_{\text{CBOA}}^{e} = H_e + \sum_\alpha\bigg(\frac12 \omega_\alpha^2 q_\alpha^2 + \omega_\alpha q_\alpha (\bm{\lambda}_\alpha \cdot \mathbf{d}) + \frac12 (\bm{\lambda}_\alpha \cdot \mathbf{d})^2 \bigg) .
\end{equation}
Potential energy surfaces (PESs) are defined by the eigenvalues $\varepsilon_k(\mathbf{R},\mathbf{q})$ of this operator
\begin{equation}
       \label{el_eq}   H_{\text{CBOA}}^{e}\psi_k({\mathbf{r};\mathbf{R},\mathbf{q}})= \varepsilon_k(\mathbf{R},\mathbf{q})\psi_k(\mathbf{r};\mathbf{R},\mathbf{q}) .
\end{equation}
Compared to the usual PESs, the modified potential energy surfaces depend on an additional set of coordinates, which are the photonic degrees of freedom. As a consequence, the interaction with the photons will lead to changes of the PESs. Note that in Eq.~\eqref{ham_cboa_orig} the non-adiabatic coupling terms have been disregarded, opposed to other proposed approaches to CBOA. \cite{cboa,fischer2023cavity}  \\ \\
\noindent
In order to solve the electronic problem defined in Eq.~\eqref{el_eq}, we need to reformulate the standard \textit{ab initio} electronic structure methods. The CBOA Hamiltonian can then be written in second quantization as
 \begin{equation}
 \label{eq:CBOA-HF-Ham}
      H_{\text{CBOA}}= \sum_{pq}\Tilde{h}_{pq}E_{pq}+\frac12\sum_{pqrs}\Tilde{g}_{pqrs}e_{pqrs} + \Tilde{c} ,
 \end{equation}
where $p,q,r,s$ denote molecular orbitals. Defining the standard one- and two- electron integrals $h_{pq}$ and $g_{pqrs}$, 
\begin{align}
    &\Tilde{h}_{pq} = h_{pq}+\sum_\alpha\big(\omega_\alpha q_\alpha\bm{\lambda}_\alpha\cdot\mathbf{d}_{pq}+\frac12\sum_s(\bm{\lambda}_\alpha\cdot\mathbf{d}_{ps})(\bm{\lambda}_\alpha\cdot\mathbf{d}_{sq}) \big)\\
    &\Tilde{g}_{pqrs} =g_{pqrs}+\sum_\alpha (\bm{\lambda}_\alpha\cdot\mathbf{d}_{pq})(\bm{\lambda}_\alpha\cdot\mathbf{d}_{rs}) \\
    \label{c_tilde}
   & \Tilde{c} = V_{NN} + \frac12\sum_\alpha \omega_\alpha^2q_\alpha^2 .
\end{align}
From Eqs.~\eqref{eq:CBOA-HF-Ham} -~\eqref{c_tilde} it follows that the implementation of the CBOA Hamiltonian only requires trivial modifications of the existing codes by means of a redefinition of the integrals. In the following, we will use these modified integrals to obtain post Hartree-Fock methods, relying on the cavity Born-Oppenheimer Hartree-Fock ansatz, already proposed in Ref. \cite{schnappinger2023cavitybornoppenheimer} 
\subsection{Comparison between the CBOA and polaritonic approaches}
\noindent
While the cavity Born-Oppenheimer approximation provides a straightforward procedure to treat the photonic degrees of freedom, this approach is not the only possible way of describing strongly coupled systems. In the past few years methods like Hartree-Fock theory, coupled cluster theory and full configuration interaction have been developed for a QED framework in the polaritonic approach. \cite{qed-cc,haugland2021intermolecular} As opposed to the CBOA, in the polaritonic framework the photonic degrees of freedom are explicitly considered in the electronic wave function, which is now more appropriately referred to as polaritonic. Here, the interactions between electrons and photons are explicitly treated and electron-photon correlation is included. On the other hand, the parametric dependence on the photonic degrees of freedom in CBOA intrinsically implies that the approach does not describe this correlation.  Along these lines, we can define the electron-photon correlation energy as the difference between the QED and the CBOA energies, calculated at the optimal value of $q$:
\begin{equation}
    E_{ep,corr} = E_{\text{QED}} - E_{\text{CBOA}, q_{opt}}
\end{equation}
As pointed out in Ref.\cite{haugland2023understanding}, the effects of electron-photon correlation on the ground state can be seen as a screening to the dipole self-energy term in the Hamiltonian. Note that in our definition of electron-photon correlation we focus on the cavity effects on the electronic wave function only. As a consequence, our definition differs from the one proposed in Ref. \cite{fischer2023cavity}, where nuclei-mediated effects are considered as well.
\\ \\
In order to analyze the potential of CBO methods, we perform a comparison of these approaches to their analogous in the polaritonic framework. Even though we expect CBO methods to be more suitable for VSC and the polaritonic approach for ESC, the possibility to tune the value of the $q$ parameter makes the cavity Born-Oppenheimer approach highly flexible. In fact, as we will show later, tuning the electromagnetic field parameter $q$ can be exploited to reproduce for instance binding energies obtained using the polaritonic approaches. Moreover, the photonic coordinates can be thought of as a proper rescaling of the expectation value of the corresponding displacement operator with respect to a photonic coherent state. With this observation in mind, CBOA can be viewed as a polaritonic mean-field approach. \\
Some similarities between the two approaches can be found at the Hartree-Fock level. In QED-HF, the energy is minimized with respect to the photonic coherent state. For this reason, CBO-HF with the value of $q$ that minimizes the energy of the system gives the same energy as QED-HF. The analytical expression for this value has been reported in Ref. \cite{schnappinger2023cavitybornoppenheimer}
\begin{equation}
    q_{opt} = -\frac{\bm{\lambda}\cdot\langle \mathbf{d}\rangle}{\omega} .
\end{equation}
With this choice for the displacement of the field, the Hamiltonians used in QED-HF and in CBO-HF are equivalent. As a consequence, calculations also inherit the origin-invariance already discussed for QED-HF,\cite{qed-cc} which would not be otherwise obtained with the CBOA Hamiltonian. Moreover, the energies do not depend on the frequency of the electromagnetic field. \cite{qed-cc}  \\ 
As far as CBO-CC is concerned, the main difference with QED-CC stands in the definition of the cluster operator. In QED-CC, the T operator also includes photonic and mixed photonic-electronic excitation operators. This method is then able to account for electron-photon correlation in the description of the polaritonic system. On the other hand, in CBO-CC only the purely electronic cluster operator is used. As a consequence, even though the effects of the electromagnetic field are parametrically taken into account, electron-photon correlation is not accounted for. For this reason, minimizing the energy for CBO-CC does not reproduce QED-CC energies when electron-photon correlation plays an important role, as in ESC regimes. In VSC regimes, on the other hand, the effect of electron-photon correlation on the ground state energy is small. As a consequence, CBO-CC for the optimal value of $q$ provides a good description of QED-CC ground state energies in VSC regimes.\cite{haugland2023understanding} Finally, fixing $q$ to minimize the electronic energy mimics what is commonly done by optimizing the nuclear geometry and thus represents a meaningful choice. As highlighted for CBO-HF, also CBO-CC results do not depend on the frequency of the field when the optimal value of $q$ is used. \cite{haugland2023understanding} 
\section{Benchmark of the methods}
\label{Sec:Benchmark}
\noindent
In order to assess the accuracy of the different methods, we consider several electronic structure methods using the Hamiltonian in Eq.~\eqref{eq:CBOA-HF-Ham}. In particular, HF, CASCI and FCI, as well as several methods in the coupled cluster hierarchy (CC2, CCSD, CC3, CCSDT) have been implemented in the CBOA framework using the $e^T$ program. \cite{eT} A detailed description of these methods can be found in the literature. \cite{pinkbook, CHRISTIANSEN1995409,cc3,ccsdt} \\
As a preliminary benchmark study, we have applied these methods to the H$_2$ dimer in parallel configuration at the optimal intermolecular distance calculated at the CCSD level. Calculations have been performed with the aug-cc-pVDZ basis set, with a fixed bond length of 0.74 $\si{\angstrom}$ for each hydrogen molecule. Only one photonic mode has been considered, with a coupling strength of $\lambda=0.1$ a.u., polarization along the intermolecular axis ($z$), and frequency 12.7 eV. The ground state energies for this system, at the optimal value of $q$, can be found in Table \ref{tab: table}.\\
\begin{table}[h!]
\centering
\begin{tabular}{lcc}
\hline
\hline
Method & Energy (eV) & Cavity effects\\
\hline
CBO-HF & -61.01990 &  0.40901 \\ 
CBO-CASCI (2,6) &  -61.03021 &  0.41213\\ 
CBO-CC2   & -62.42728 & 0.49271\\ 
CBO-CCSD & -63.00968 & 0.37156\\
CBO-CC3 & -63.00984 & 0.37163\\ 
CBO-CCSDT  & -63.00991 & 0.37163\\    
CBO-FCI &  -63.00991 & 0.37163\\	
\hline
\hline
\end{tabular}
\caption{H$_2$ - H$_2$ ground state energy with different CBO methods. For each method, cavity effects are defined as $\text{E}_{\text{CBO}} - \text{E}_{\text{no cavity}}$.}
\label{tab: table}
\end{table}
From this Table, we note that the introduction of electronic correlation in post CBO-HF methods significantly increases the accuracy of the results. Nevertheless, CBO-CC2 overestimates the effects of the electromagnetic field on this system. Overall, CBO-CCSD is able to provide an accurate description of electronic correlation. For this reason, only CBO-CCSD will be considered in the following discussions.
\section{Intermolecular interactions in dimers}
\label{Sec:Intermolecular interactions in dimers}
\noindent 
In order to assess the usefulness of CBO-CCSD, we have performed calculations of the intermolecular interactions in dimer complexes. Despite the usual weak character of intermolecular interactions, they play an essential role in chemistry, since they are critical to the supramolecular organization of every system.
Moreover, experimental results have shown the possibility to exploit resonant cavities to obtain different supramolecular organizations, \cite{joseph2021supramolecular,nagarajan2021chemistry,hirai2021selective} thus highlighting how strong light-matter coupling can affect non-covalent interactions between molecules. This phenomenon is probably strongly related to the cooperative effects between molecules and is quite complicated to analyze theoretically. In fact, the description requires accurate \textit{ab initio} methods that are able to treat a large number of molecules with reasonable computational cost. Nevertheless, a benchmark study of intermolecular interactions in dimers has recently been presented using QED-CCSD, \cite{haugland2021intermolecular} where the cavity-induced effects are explicitly taken into account in the calculation of polaritonic potential energy surfaces. \\
In the following, a different point of view will be provided using the cavity Born-Oppenheimer approximation. Even though this method is not able to explicitly take into account the polaritonic character of potential energy surfaces, it provides a good representation of the ground state properties. In the CBO-CCSD calculations, only one photonic mode has been considered with coupling strength $\lambda =0.1$ a.u.. This relatively large coupling mimics the use of several modes.  All the presented potential energy surfaces are relative to the monomers at a distance of 200 $\si{\angstrom}$. Where possible, the results are compared to the polaritonic approach in Ref. \cite{haugland2021intermolecular}
\subsection{Hydrogen dimer}
\label{Sec:Hydrogen dimer}
\begin{figure}[htp]%
    \centering%
    \includegraphics[width=\columnwidth]{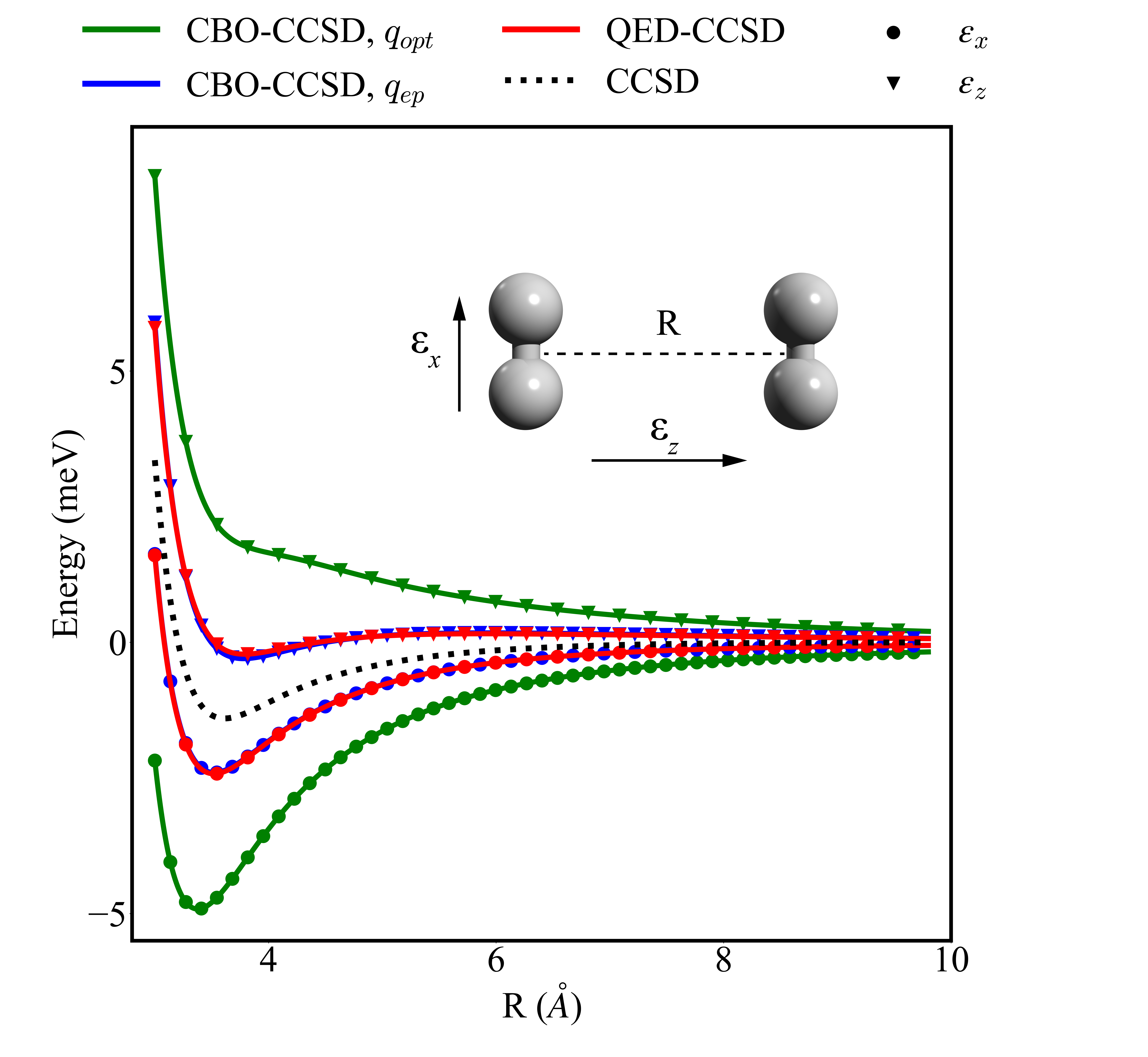}%
    \caption{Potential energy curves the H$_2$ dimer in parallel configuration. $\lambda=0.1$ a.u., $\omega=12.7$ eV. The field polarization is along the intermolecular axis ($z$) or the H$_2$ axis ($x$). The energy at 200 \si{\angstrom} has been subtracted.}%
    \label{fig:2-H2_par-x_z}%
\end{figure}

\begin{figure*}%
\centering%
    \includegraphics[width=\textwidth]{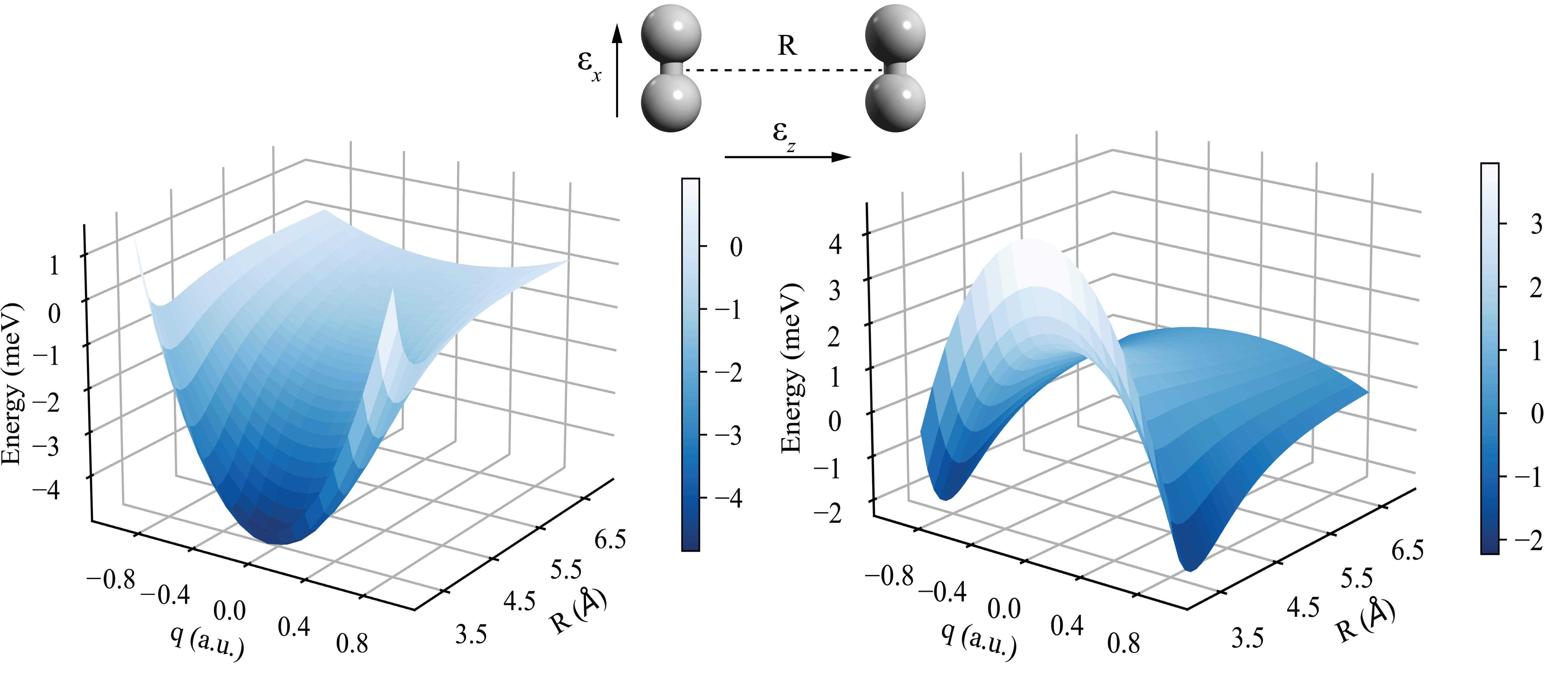}%
    \caption{CBO-CCSD potential energy surfaces for the H$_2$ dimer in the parallel configuration. $\lambda=0.1$ a.u., $\omega=12.7$ eV. The field polarization is along the H$_2$ axis ($x$, left) or the intermolecular axis ($z$, right). At each $q$, the energy at 200 \si{\angstrom} has been subtracted.}
    \label{fig:2-H2_par-3d_z}%
\end{figure*}

\begin{figure*}[]%
    \centering%
    \includegraphics[width=\textwidth]{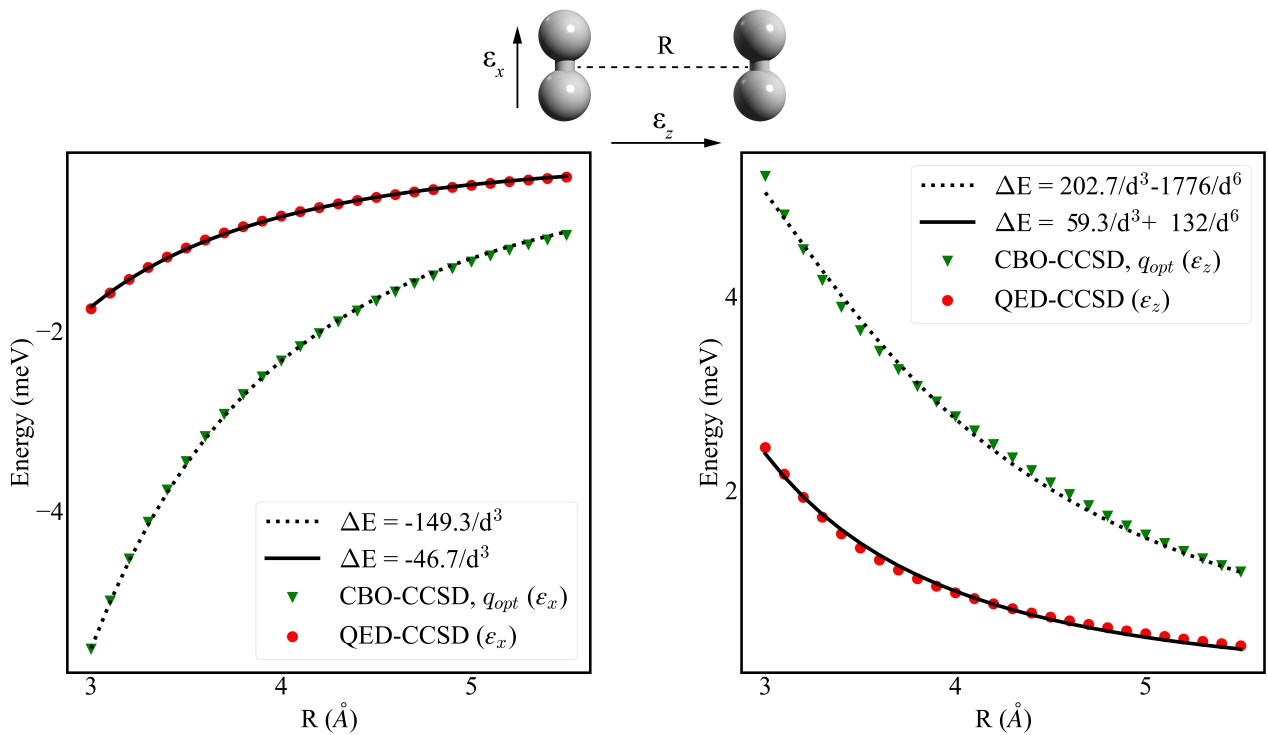}%
    \caption{Cavity-induced effects on the CBO-CCSD potential energy curves (E$_{\text{CBO-CCSD}}$-E$_{\text{CCSD}}$; E$_{\text{QED-CCSD}}$-E$_{\text{CCSD}}$) for the H$_2$ dimer in the parallel configuration. $\lambda=0.1$ a.u., $\omega=12.7$ eV. The field polarization is along the H$_2$ axis ($x$, left) or the intermolecular axis ($z$, right). The energy at 200 \si{\angstrom} has been subtracted. All fitted curves have $R^2>0.996$.}%
    \label{fig:2-H2_par-dipoli}%
\end{figure*}

\noindent
We consider the hydrogen dimer in parallel configuration (See Fig.~\ref{fig:2-H2_par-x_z}). Outside the cavity, the interactions of the hydrogen molecules are dominated by weak dispersion interactions, closely related to electron correlation. The interaction energy thus scales as $R^{-6}$ and can be expressed in terms of the polarizability of the monomers.\cite{atkins2011molecular} \\
The potential energy surfaces have been calculated using CBO-CCSD by varying both the intermolecular distance and the photonic coordinate using the aug-cc-pVDZ basis set. As in Sec.~\ref{Sec:Benchmark}, the H-H distance of each monomer has been kept fixed at the equilibrium value of 0.74 $\si{\angstrom}$ and the electromagnetic field frequency was set to 12.7 eV, in resonance with the first electronic excitation of the system calculated at the CCSD level. In this case, two different field polarizations have been considered: the intermolecular axis ($z$) and the H$_2$ bond axis ($x$).\\ \\
In Fig.~\ref{fig:2-H2_par-x_z}, we present CBO-CCSD potential energy surfaces for different values of $q$, together with the QED-CCSD and the CCSD curves. We show curves for the optimal value of $q$ ($q_{opt}$) and for the values of $q$ that make CBO-CCSD as close as possible to QED-CCSD ($q_{ep}$). 
In Fig.~\ref{fig:2-H2_par-3d_z}, instead, the complete three-dimensional potential energy surfaces for both polarizations are reported. 
We note that the field polarization along $z$ causes a destabilization of the system and the binding energy increases when the polarization is along $x$. \\
For CBO-CCSD, $q_{opt}=0$ for both polarization directions. In particular, when the field polarization is along $z$, the potential energy curve is repulsive at every intermolecular distance. However, when increasing the absolute value of $q$ the system becomes more stabilized, as can also be seen from Fig.~\ref{fig:2-H2_par-3d_z} (\textit{right}). For the $x$ polarization, CBO-CCSD shows a considerable stabilization of the system at $q_{opt}=0$, while the increase of this parameter leads to weaker intermolecular interactions (See Fig.~\ref{fig:2-H2_par-3d_z}, \textit{left}). We also notice from Fig.~\ref{fig:2-H2_par-3d_z} that the binding energies are symmetric with respect to $q\rightarrow -q$ for this geometry of the system.\\
For $q=0$, besides minimizing the energy of the system, this also leads to a Hamiltonian that only contains the dipole self-energy term. Increasing the value of $q$ allows us to tune the relative importance of this term with respect to the bilinear one. We now focus on the values of $q$ that closely mimic the QED-CCSD binding energies. These values are very similar for both polarization directions ($q_{ep}^x=0.655$ and $q_{ep}^z=0.666$), and have been determined by a manual scan of the $q$ parameter. Overall, we conclude that cavity Born-Oppenheimer methods are not able to describe electron-photon correlation, but we can mimic this correlation by varying the value of $q$. \\ \\
\noindent
In order to rationalize the trends observed so far, it is interesting to consider cavity-induced effects on the binding energy. In Fig.~\ref{fig:2-H2_par-dipoli} we analyze the $x$ and $z$ polarizations and compare CBO-CCSD for the optimal value of $q$ with QED-CCSD. We note that the $R^{-3}$ scaling of QED-CCSD is reproduced by CBO-CCSD. However, the coefficients found with the two approaches differ considerably. This further highlights the lack of electron-photon correlation in CBO-CCSD. These observations hold for both polarizations considered.
\subsection{Water dimer}
\begin{figure}[htp]%
    \centering%
    \includegraphics[width=\columnwidth]{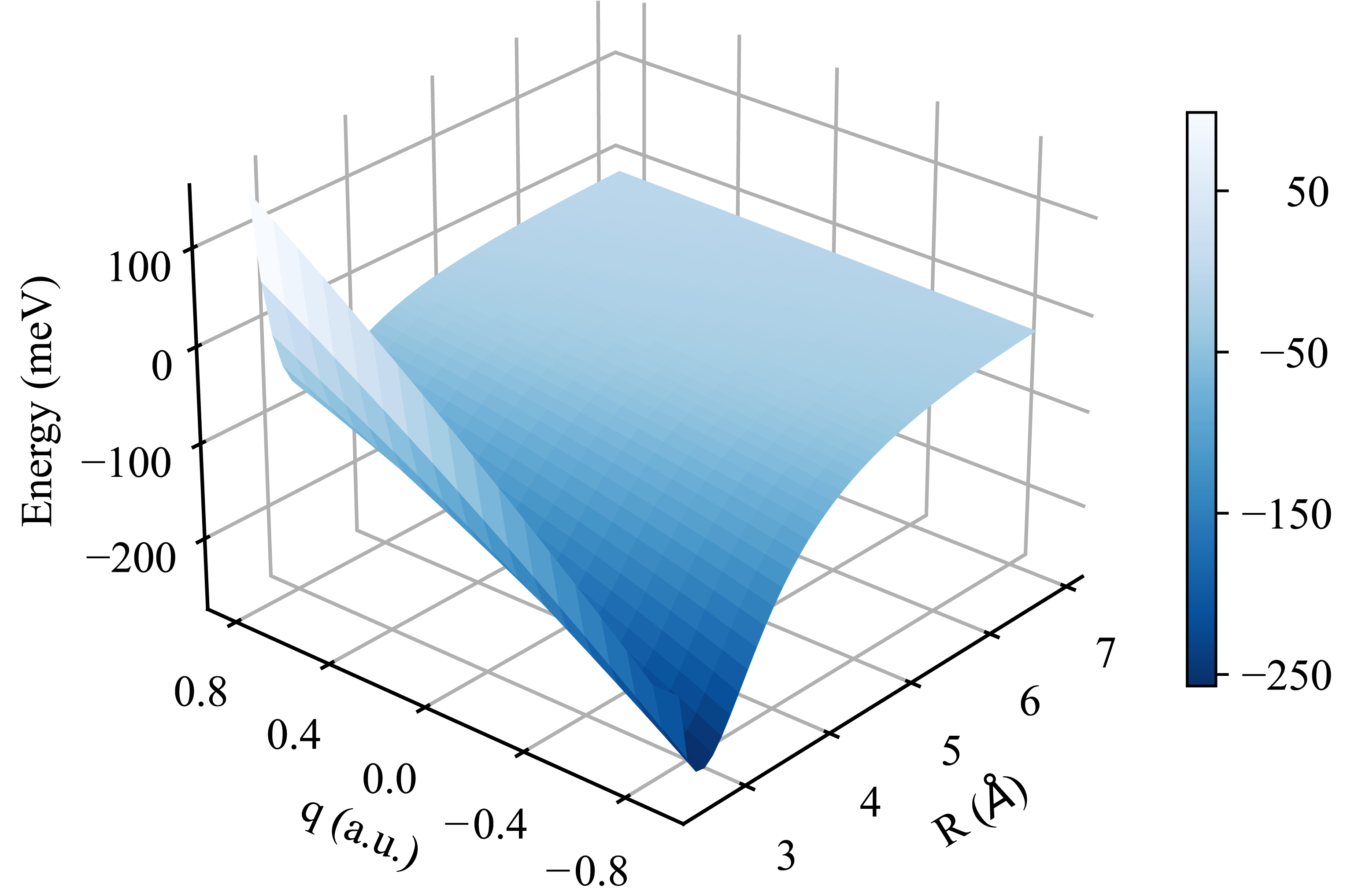}%
    \caption{Potential energy surface for two parallel H$_2$O molecules, $\lambda=0.1$, $\omega=7.86$ eV. The field polarization is along the chain axis ($z$). For each $q$, the energy at 200.0 \si{\angstrom} has been subtracted.}%
    \label{fig:2-H2O-3d}%
\end{figure}

\noindent
The second system considered is the water dimer. In this case, intermolecular interactions are more pronounced and characterized by a strong hydrogen bond. The main contribution to this bond is provided by dipole-dipole interactions, even though a consistent component of charge transfer is also participating.\cite{ronca2014quantitative,cappelletti2012revealing} The study of such a system is of great interest, since water plays an important role in several chemical phenomena. The possibility to modulate hydrogen bonding would be an important tool in several fields of application, ranging from catalysis to biological phenomena. \\
We report calculations at different values of $q$ varying the distance between the two oxygen atoms. The orientations of the water molecules are kept fixed for every intermolecular distance, as well as the geometry of each monomer, which we obtained from Ref. \cite{ronca2014quantitative} The frequency of the electromagnetic field is 7.86 eV, resonant with the first electronic excitation calculated at the CCSD level. The field polarization is parallel to the axis between the oxygen atoms ($z$). \\ 
\begin{figure}[htp]%
    \centering%
    \includegraphics[width=\columnwidth]{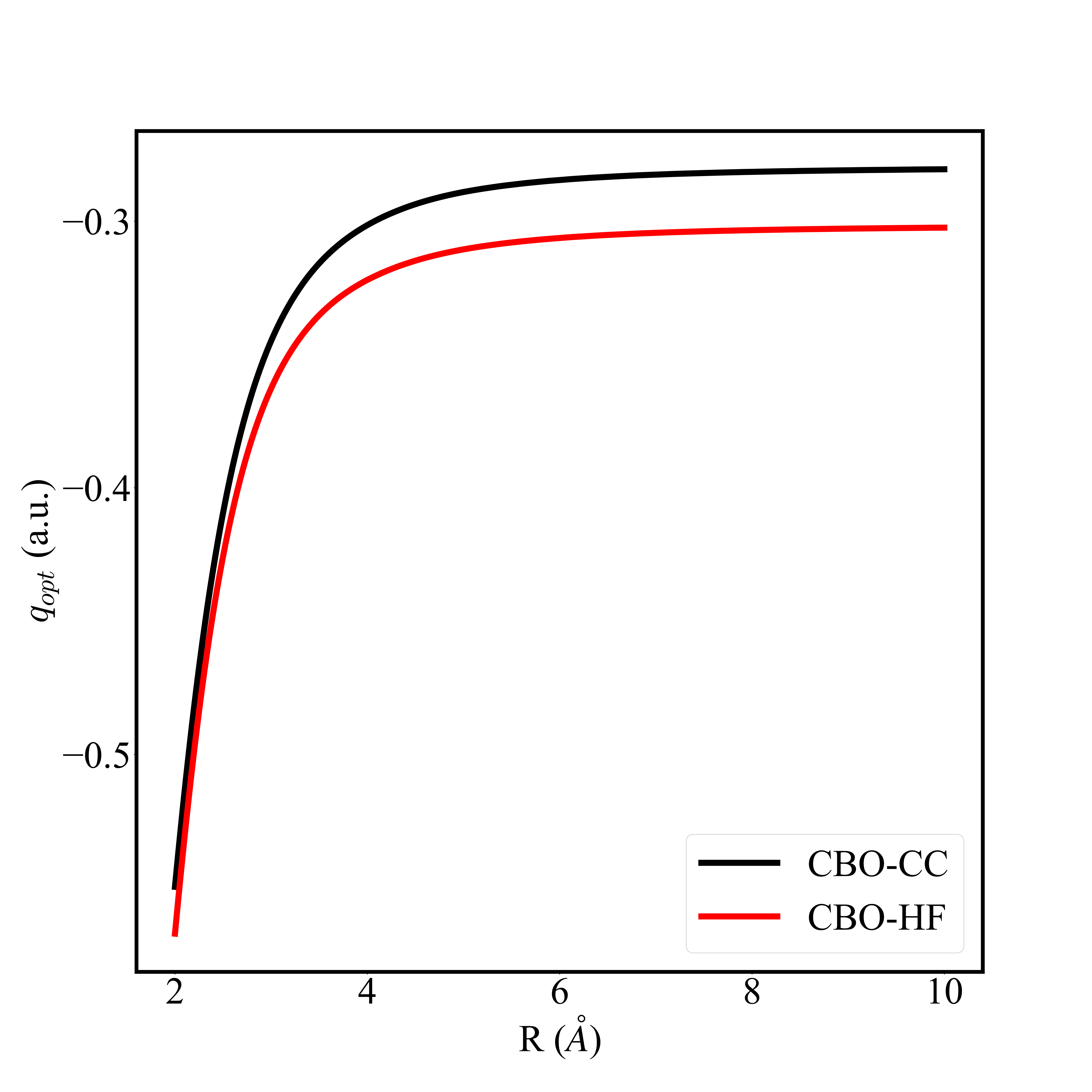}%
    \caption{Trend of the optimal value of $q$ with respect to the intermolecular distance.}%
    \label{fig:2-H2O-opt_q}%
\end{figure}%
\noindent
We determined the optimal value of $q$ by minimizing the total energy of the system at every intermolecular distance $R$. The dependence of $q_{opt}$ with respect to $R$ is shown in Fig.~\ref{fig:2-H2O-opt_q} and displays an interesting behavior of the system. When increasing the distance between the molecules, the dimer requires a smaller absolute value of $q$ to be stabilized. Both CBO-HF and CBO-CCSD show this trend, although for CBO-HF the absolute value of $q$ is larger. This suggests that when explicitly including electron correlation the system requires smaller absolute values of the displacement of the field to be stabilized. \\ 
\begin{figure}[htp]%
    \centering%
    \includegraphics[width=\columnwidth]{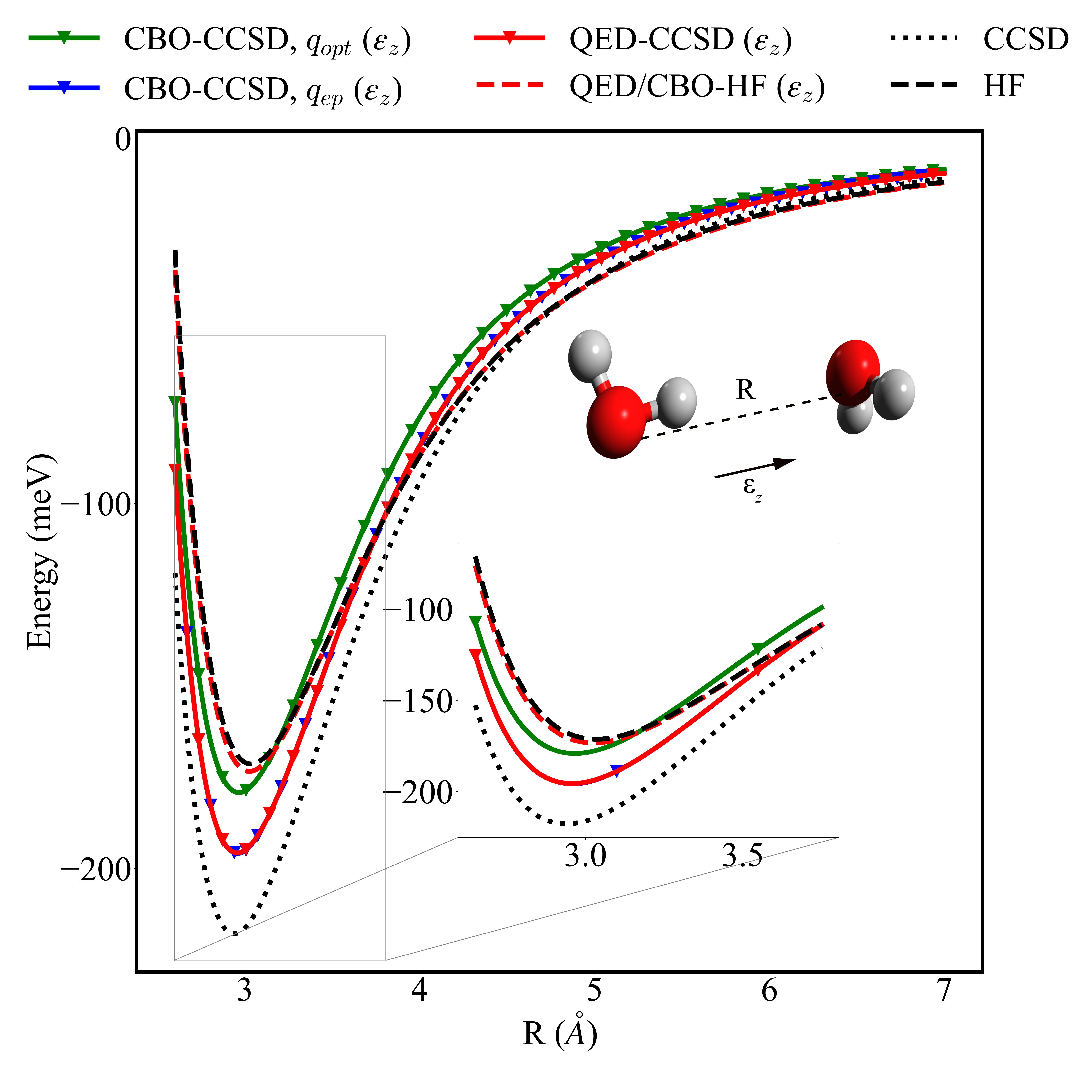}%
    \caption{Potential energy curves with different electronic structure methods for the water dimer, $\lambda=0.1$ a.u., $\omega=7.86$ eV. The field polarization is along the intermolecular axis ($z$). The energy at 200.0 \si{\angstrom} has been subtracted.}%
    \label{fig:2-H2O}%
\end{figure}%
\noindent
We now turn to the potential energy curves reported in Fig.~\ref{fig:2-H2O} for CBO-CCSD at  $q_{opt}$ and $q_{ep}$, together with QED-CCSD and CCSD. We also report results at the Hartree-Fock level, and we only show one curve for QED-HF and CBO-HF since the two methods are equivalent. \\
From Fig.~\ref{fig:2-H2O}, we observe that the system is destabilized by the electromagnetic field at the CCSD level. The CBOA approach with the optimal value of $q$ clearly underestimates the binding energy inside the cavity. When increasing the displacement field the system is further destabilized, as can be observed from Fig.~\ref{fig:2-H2O-3d}. The binding energies do not show any particular symmetry with respect to the variation of the $q$ parameter (See Fig.~\ref{fig:2-H2O-3d}). Even though a proper interpretation of this behavior would require further investigation, this observation seems to be connected to the low symmetry of the system, as the dipole moment is changed considerably by variations in the electromagnetic field. However, it is possible to choose $q$ such that CBO-CCSD closely mimics QED-CCSD. Depending on the geometry of the system, the values of $q_{ep}$ range from $-0.47$ to $-0.42$. \\
Finally, from Fig.~\ref{fig:2-H2O} we can also analyze the effects of electron and electron-photon correlation. Outside the cavity, electron correlation leads to a considerably larger binding energy. Inside the cavity, the CBOA overestimates the effects of the field and leads to a smaller binding energy compared to QED-CCSD. The difference between CBO-CCSD and QED-CCSD is due to electron-photon correlation, which is not captured in the cavity Born-Oppenheimer approach.\cite{haugland2023understanding}  Nevertheless, CBO-CCSD calculations with the optimal value of $q$ are still able to partially capture the cavity-induced destabilization of the system.\\
\subsection{Benzene-water}
\begin{figure}[htp]%
   \centering%
    \includegraphics[width=\columnwidth]{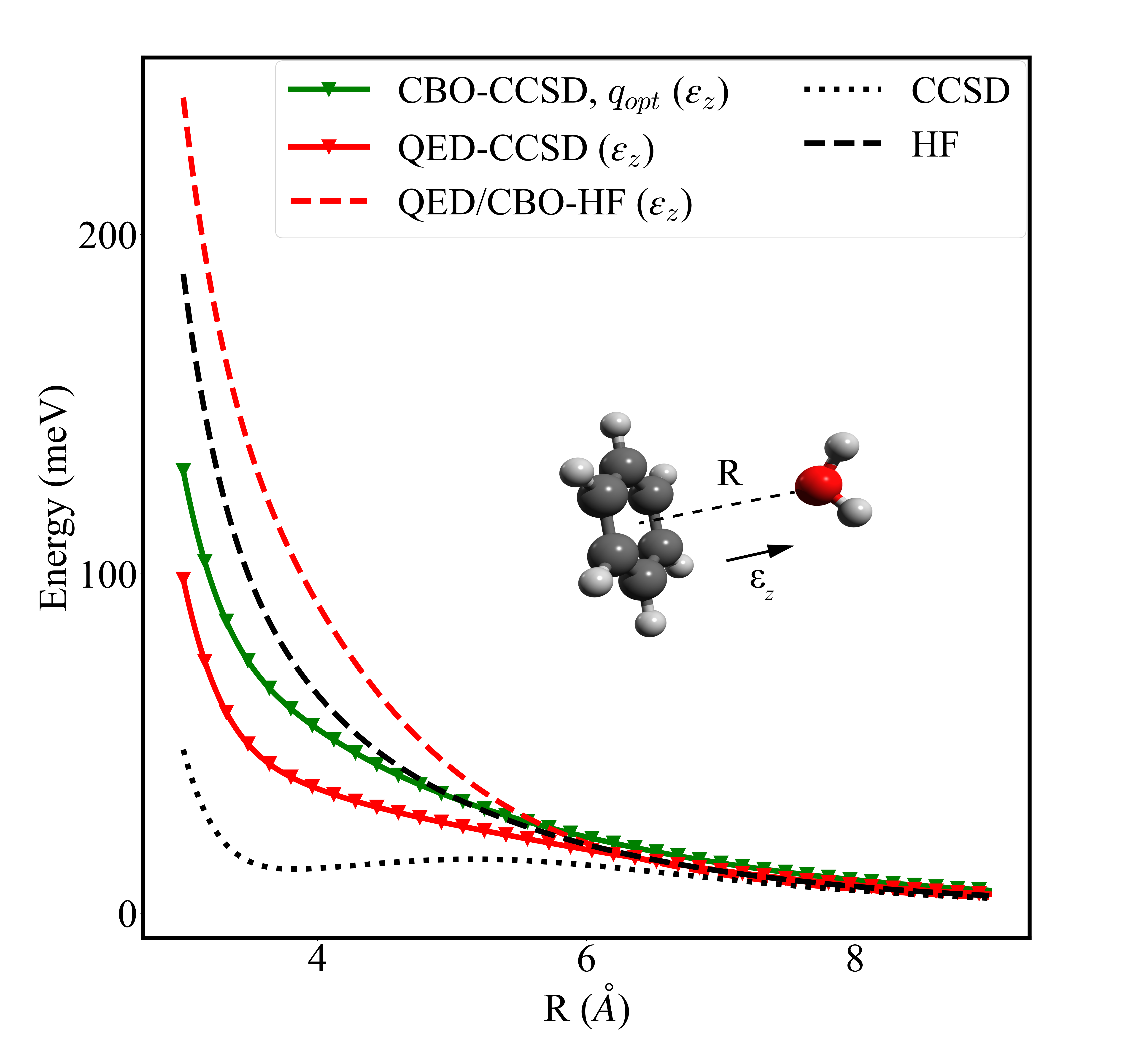}%
    \caption{Potential energy surface for the benzene-water system, $\lambda=0.1$, $\omega$=13.6 eV. The field polarization is along the intermolecular axis ($z$). The energy at 200.0 \si{\angstrom} has been subtracted.}%
   \label{fig:h2o_benz}%
\end{figure}

\noindent
We now consider the interaction between benzene and water. In this dimer, the polar character of water induces charge fluctuations in benzene, and intermolecular interactions are usually described in terms of dipole-induced dipole forces.\cite{atkins2011molecular}\\
We report binding energies for this complex studied by varying the distance between the oxygen atom of the water molecule and the benzene ring. The relative orientation of the single monomers is kept fixed, with the oxygen pointing towards the ring (See Fig.~\ref{fig:h2o_benz}). The field polarization is parallel to the dipole moment of the system and the frequency is 13.6 eV. We used the 6-31+G* basis set.\\
In Fig.~\ref{fig:h2o_benz} we show the potential energy curves outside and inside the cavity, with the polaritonic and the CBOA approaches. \\
We note that the electromagnetic field clearly destabilizes the system for all approaches. Furthermore, we see that electron correlation plays an important role in the stabilization of this system outside the cavity. Inside the cavity, CBOA reproduces this trend to a smaller extent. Finally, for QED-CCSD electron-photon correlation further stabilizes the system, as is the case for the water dimer system.
\section{Conclusions}
\label{Sec:Conclusions}
\noindent
In this work, we have presented an implementation of the cavity Born-Oppenheimer approximation for several electronic structure methods. In particular, this framework enables a way to parametrically introduce the effects of the electromagnetic field on the electronic wave functions and potential energy surfaces. We have benchmarked CBOA methods against the QED-CCSD reference for intermolecular interactions in dimers. In particular, we have focused on van der Waals complexes, hydrogen bonds and dipole-induced dipole interactions. Such a comparison can be particularly useful to understand the contribution of electron-photon correlation to polaritonic potential energy surfaces. Moreover, it has been shown that the $R^{-3}$ scaling of the van der Waals interactions found at the QED-CCSD level \cite{haugland2021intermolecular} is reproduced by CBO-CCSD as well. \\
The cavity Born-Oppenheimer approach provides a more cost-effective way of describing the electromagnetic field, although at a lower accuracy. The main component missing in the CBOA approach is the explicit treatment of electron-photon correlation. As a consequence, we expect CBO methods to perform well in the vibrational strong coupling regime. For electronic strong coupling conditions, instead, these methods only provide an approximate description. Nevertheless, the approach could be particularly useful in \textit{ab initio} molecular dynamics simulations of strongly coupled systems, due to the reduced computational cost and the straightforward implementation in existing electronic structure codes.
\begin{acknowledgments}
\noindent
We acknowledge Matteo Rinaldi for the implementation of the CASCI method. S.A. and H.K. acknowledge funding from the European Research Council (ERC) under the European Union's Horizon 2020 Research and Innovation Programme (grant agreement No. 101020016). T.S.H. and H.K. acknowledge funding from the  Research Council of Norway through FRINATEK project 275506. E.R. acknowledges funding from the European Research Council (ERC) under the European Union’s Horizon Europe Research and Innovation Programme (Grant n.ERC-StG-2021-101040197-QED-SPIN).
\end{acknowledgments}
\bibliography{bibliography}
\end{document}